# The Photometric Period and Variability of the Cataclysmic Variable V849 Herculis (PG 1633+115)

F. A. Ringwald*, Gerald D. Rude II, and Jonathan J. Roveto
Department of Physics
California State University, Fresno
2345 E. San Ramon Ave., M/S MH37
Fresno, CA 93740-8031, U. S. A.

Kelly S. Khamvongsa
Research Experiences for Undergraduates Program
Department of Astronomy
San Diego State University
5500 Campanile Drive, PA-210
San Diego, CA 92182-1221, U. S. A.

Also:
Department of Mathematics
California State University, Fresno
5245 N. Backer Ave., M/S PB108
Fresno, CA 93740-8001, U. S. A.

ABSTRACT

We report time-resolved photometry of the cataclysmic variable V849 Her, and measure a period of 0.1414 ± 0.0030 days (3.394 ± 0.072 hours). We also present photometry taken over several weeks in 2010 and 2011, as well as light curves from 1995 to 2011 by the American Association of Variable Star Observers. The spectra, absolute magnitude derived from infrared magnitudes, and variability all suggest that V849 Her is a nova-like variable. The shallow (0.5-magnitude) low states we observe resemble the erratic low states of the VY Sculptoris stars, although they may recur quasi-periodically over an average cycle of 12.462 ± 0.074 days.

* Corresponding author. Tel: +1-559-278-8426; fax: +1-559-278-7741.
*E-mail address*: ringwald@csufresno.edu

## 1. Introduction

V849 Herculis was discovered in the Palomar-Green survey by Green et al. (1986). They listed it as PG 1633+115, gave a finding chart, and classified it as COMP: CV:. The colons denoted uncertainty: this meant either a composite-spectrum object, likely a binary star system "in which the blue component dominates the spectrum shortward of 5000 Å," or a cataclysmic variable star system, with "emission lines at velocities that place them within the Galaxy." They listed an apparent photographic magnitude $B_{pg}$ = 14.92. PG 1633+115 was on the list of DA white dwarfs of Fleming et al. (1986). Beers et al. (1992) gave $V$ = 15.3 and classified it as a metal-poor sdB star, with weak Ca II H and K lines. As this paper will show, it is certainly not a single white dwarf, or a hot subdwarf.

Misselt and Shafter (1995) obtained time-resolved photometry, through both Johnson $V$ and Cousins $R$ filters. They showed that PG 1633+115 is definitely a cataclysmic variable, since it shows the erratic flickering characteristic of cataclysmic variables (Chapter 10 of Hellier, 2001). Misselt and Shafter also found a sinusoidal modulation in apparent magnitude, with a low amplitude $\Delta V \sim 0.06$. They were unable to measure the precise period of this variability, because of aliasing (page 131 of Hellier, 2001; see also Thorstensen and Freed, 1985). The most likely photometric periods found by Misselt and Shafter were 3.94 hours and 3.38 hours, which are one cycle/day aliases of each other, and 3.43 hours, an alias of the 3.38-hour periodicity.

Misselt and Shafter (1995) also found that PG 1633+115 can vary by as much as $\Delta V \sim$ 0.3 in 10 days. This could be from dwarf nova outbursts (Chapter 5 of Hellier, 2001), although it might alternatively be from low states, as in VY Scl and similar stars (see Chapter 12 of Hellier, 2001). PG 1633+115 was given the variable star name V849 Herculis by Kazarovets and Samus (1997), who listed $V_{max}$ = 15.0 and a variability of 0.5 magnitudes, on the basis of the observations of Misselt and Shafter.

Rodríguez-Gil et al. (2007) did a radial-velocity study of V849 Her. They measured a spectroscopic period of 3.15 ± 0.48 hours, based on only 2.64 hours of observations made during one night. Spectroscopic periods generally do measure the orbital periods more reliably than photometric periods, save for in eclipsing cataclysmic variables, but Rodríguez-Gil et al. admitted that the orbital period of V849 Her is "not a settled issue."

## 2. Observations

*2.1 Spectra*

The blue spectrum shown in Figure A1 in Appendix A (electronically available only), and also shown by Ringwald (1993), shows a hint of possible weak Hα emission, and weak, broad Hβ absorption. A red spectrum, taken six days later and shown in Figure A2 in Apprendix A, shows weak, broad Hα emission on a noticeably fainter continuum. Both spectra were taken with the Hiltner 2.4-m telescope, Mark III spectrograph, and BRICC camera at MDM Observatory (Luppino, 1989), and have 11-Å resolution, at a dispersion of 5 Å/pixel. The emission lines and continuum variability in these spectra confirm that V849 Her is a cataclysmic variable, and suggest that it has a high mass-transfer rate, similar to a dwarf nova in outburst, since the continuum is so strong and the lines are so weak (see Figure 3.6 of Hellier, 2001).

Munari and Zwitter (1998) and Rodríguez-Gil et al. (2007) both showed spectra similar to that in Figure A1, with weak Hα emission and broad Hβ absorption on a bright, blue continuum that can vary over days. That all the available spectra show weak lines on a bright, blue continuum suggests that V849 Her is in a high state most of the time, or in other words is a nova-like variable. A nova-like resembles a dwarf nova stuck in outburst, because it has a high mass-transfer rate through its accretion disk (see page 72 of Hellier, 2001).

*2.2 Photometry*

We obtained time-resolved, differential photometry of V849 Her, both with Fresno State's station at Sierra Remote Observatories (SRO), and at Mount Laguna Observatory (MLO). The Sierra Remote observations used the station's 16-inch f/8 DFM Engineering telescope, its Santa Barbara Instruments Group STL-11000M CCD camera, and a clear Astrodon filter.

For the SRO photometry, exposure times were 120 seconds. The CCD was binned 3x3 to reduce the dead time between exposures to 7 seconds, making for a total of about 127 seconds per exposure. The CCD had a temperature of −5º C, so each night 9-15 dark frames of the same exposure times as the target frames were taken, median combined, and subtracted from all target frames, to reduce thermal noise from the camera.

The image scale for all SRO photometry was 1.71 arcseconds/pixel, since the detector was binned 3x3. This image scale was almost always larger than the seeing, which is typically 1-1.5 arcseconds at this site. Exposure times were precise only to within 1 second, since all timing was done with the clock of the personal computer used to run the STL-11000M camera. This computer clock was synchronized at the beginning of each night with the time-signal clock of the National Institute of Standards and Time (NIST) over the Internet, with their free program NISTime 32 (Levine, 2010).

The observations at Mount Laguna Observatory used its 1-m telescope, its CCD 2005 camera, and no filter. Exposure times were 5 seconds, with 2.31 seconds of dead time between exposures, since the CCD 2005 detector was binned 4x4. The CCD 2005 camera is liquid-nitrogen cooled, with negligible thermal noise. The image scale was 1.62 arcseconds/pixel for the Mount Laguna photometry, since the detector was binned 4x4. Time for the Mount Laguna photometry was again kept only by a computer clock signal, so it should be trusted only to within 1 second.

All data reductions used AIP4WINv2 software (Berry and Burnell, 2005). Magnitudes were measured with aperture photometry (see Chapter 10 of Berry and Burnell, 2005), with an aperture radius of 6 pixels centered on the star, with sky background measured inside an annulus between 9 and 12 pixels around the star. The comparison star (C1) used to measure the photometry was GSC 968:966, and the check star (C2) was GSC 969:1956. No flat-field calibrations were done, but the standard deviation of the C2-C1 magnitudes for all 2011 observations was 1.73%, so the photometry should be accurate to within 0.02 magnitudes. Table 1 is a journal of observations.

**Table 1:** Journal of observations

| UT date | UT Start | Filter | Duration (hours) | Time Resolution (seconds) | Instrument |
|---|---|---|---|---|---|
| 2010 June 06 | 04:31 | Clear | 7.02 | 120+7 | SRO 16″/STL-11K |
| 2010 June 07 | 04:22 | Clear | 7.30 | 120+7 | SRO 16″/STL-11K |
| 2010 June 10 | 05:00 | Clear | 6.61 | 120+7 | SRO 16″/STL-11K |
| 2010 June 12 | 04:20 | Clear | 7.27 | 120+7 | SRO 16″/STL-11K |
| 2010 June 15 | 04:13 | Clear | 7.12 | 120+7 | SRO 16″/STL-11K |
| 2010 June 16 | 04:17 | Clear | 7.01 | 120+7 | SRO 16″/STL-11K |
| 2010 June 19 | 04:23 | None | 6.32 | 5+2.31 | MLO 1m/CCD2005 |
| 2010 June 20 | 03:45 | None | 6.79 | 5+2.31 | MLO 1m/CCD2005 |
| 2011 June 22 | 05:30 | Clear | 5.22 | 120+7 | SRO 16″/STL-11K |
| 2011 June 23 | 04:26 | Clear | 6.06 | 120+7 | SRO 16″/STL-11K |
| 2011 June 24 | 04:17 | Clear | 6.34 | 120+7 | SRO 16″/STL-11K |
| 2011 June 25 | 04:40 | Clear | 5.75 | 120+7 | SRO 16″/STL-11K |
| 2011 June 26 | 05:02 | Clear | 5.59 | 120+7 | SRO 16″/STL-11K |
| 2011 June 27 | 04:49 | Clear | 5.49 | 120+7 | SRO 16″/STL-11K |
| 2011 June 28 | 04:28 | Clear | 5.95 | 120+7 | SRO 16″/STL-11K |
| 2011 July 03 | 04:31 | Clear | 5.48 | 120+7 | SRO 16″/STL-11K |
| 2011 July 04 | 05:14 | Clear | 4.63 | 120+7 | SRO 16″/STL-11K |
| 2011 July 09 | 06:04 | Clear | 3.53 | 120+7 | SRO 16″/STL-11K |
| 2011 July 10 | 04:18 | Clear | 5.24 | 120+7 | SRO 16″/STL-11K |
| 2011 July 11 | 04:26 | Clear | 4.99 | 120+7 | SRO 16″/STL-11K |
| 2011 July 12 | 05:26 | Clear | 4.10 | 120+7 | SRO 16″/STL-11K |
| 2011 July 13 | 05:46 | Clear | 3.74 | 120+7 | SRO 16″/STL-11K |
| 2011 July 17 | 04:42 | Clear | 4.56 | 120+7 | SRO 16″/STL-11K |
| 2011 July 18 | 04:08 | Clear | 5.03 | 120+7 | SRO 16″/STL-11K |
| 2011 July 19 | 04:54 | Clear | 4.24 | 120+7 | SRO 16″/STL-11K |
| 2011 July 20 | 05:17 | Clear | 3.46 | 120+7 | SRO 16″/STL-11K |
| 2011 July 21 | 04:41 | Clear | 4.02 | 120+7 | SRO 16″/STL-11K |
| 2011 July 22 | 04:10 | Clear | 4.45 | 120+7 | SRO 16″/STL-11K |
| 2011 July 23 | 04:52 | Clear | 4.00 | 120+7 | SRO 16″/STL-11K |
| 2011 July 24 | 04:11 | Clear | 4.32 | 120+7 | SRO 16″/STL-11K |
| 2011 July 25 | 04:10 | Clear | 4.56 | 120+7 | SRO 16″/STL-11K |
| 2011 July 26 | 04:12 | Clear | 4.43 | 120+7 | SRO 16″/STL-11K |
| 2011 July 27 | 04:44 | Clear | 3.88 | 120+7 | SRO 16″/STL-11K |
| 2011 July 28 | 04:03 | Clear | 4.23 | 120+7 | SRO 16″/STL-11K |
| 2011 July 29 | 05:51 | Clear | 2.60 | 120+7 | SRO 16″/STL-11K |
| 2011 July 31 | 05:42 | Clear | 2.49 | 120+7 | SRO 16″/STL-11K |
| 2011 August 01 | 05:18 | Clear | 2.14 | 120+7 | SRO 16″/STL-11K |
| 2011 August 02 | 03:57 | Clear | 4.00 | 120+7 | SRO 16″/STL-11K |
| 2011 August 03 | 04:09 | Clear | 1.07 | 120+7 | SRO 16″/STL-11K |
| 2011 August 05 | 04:02 | Clear | 1.11 | 120+7 | SRO 16″/STL-11K |
| 2011 August 06 | 04:02 | Clear | 1.14 | 120+7 | SRO 16″/STL-11K |
| 2011 August 07 | 04:06 | Clear | 1.03 | 120+7 | SRO 16″/STL-11K |
| 2011 August 08 | 04:29 | Clear | 0.89 | 120+7 | SRO 16″/STL-11K |
| 2011 August 10 | 03:55 | Clear | 1.38 | 120+7 | SRO 16″/STL-11K |
| 2011 August 11 | 05:04 | Clear | 1.39 | 120+7 | SRO 16″/STL-11K |
| 2011 August 12 | 03:51 | Clear | 1.05 | 120+7 | SRO 16″/STL-11K |
| 2011 August 14 | 03:58 | Clear | 0.93 | 120+7 | SRO 16″/STL-11K |
| 2011 August 16 | 03:50 | Clear | 1.01 | 120+7 | SRO 16″/STL-11K |
| 2011 August 19 | 03:42 | Clear | 1.21 | 120+7 | SRO 16″/STL-11K |
| 2011 August 20 | 03:41 | Clear | 1.24 | 120+7 | SRO 16″/STL-11K |
| 2011 August 21 | 03:37 | Clear | 1.00 | 120+7 | SRO 16″/STL-11K |
| 2011 August 23 | 04:01 | Clear | 1.00 | 120+7 | SRO 16″/STL-11K |
| 2011 August 25 | 04:40 | Clear | 1.78 | 120+7 | SRO 16″/STL-11K |
| 2011 August 26 | 05:32 | Clear | 0.50 | 120+7 | SRO 16″/STL-11K |
| 2011 August 28 | 05:23 | Clear | 0.96 | 120+7 | SRO 16″/STL-11K |
| 2011 August 29 | 04:01 | Clear | 1.04 | 120+7 | SRO 16″/STL-11K |

Note to Table 1: The time resolution column lists exposure times (e.g. 120 seconds for the first night) + dead time between exposures, in which the CCD was being read out (e.g. 7 seconds dead time between exposures, for the SRO data).

*2.3 Analysis of light curves*

Figure 1 shows each night's light curves from 2010, offset so the plots don't overlap. All show the erratic flickering characteristic of cataclysmic variables, thought to be from their accretion disks (Chapter 10 of Hellier, 2001). Figure 2 is a Lomb-Scargle periodogram (Lomb, 1976; Scargle, 1982; Chapter 13.8 of Press et al., 2007) of the SRO photometry for 2010 June 15 and 16 UT. It shows a periodicity of 7.073 ± 0.150 cycles/day, which corresponds to a period of 0.1414 ± 0.0030 days (3.394 ± 0.072 hours), as well as a surrounding forest of one cycle/day aliases. This signal is also present during other nights (see Figure 1), and in our 2011 photometry. The peak in Figure 2 at 2.166 cycles/day and other signals at frequencies lower than 4 cycles/day (corresponding to periods of 6 hours or longer) were not reproduced during other nights. We therefore interpret these signals to have been low-frequency noise, caused by the day-to-day variability of the light curve of V849 Her, shown in Figure 3.

Figure A3 in Appendix A is a log-log plot of a Lomb-Scargle periodogram of the photometry taken at Mount Laguna through no filter on 2010 June 19 and 20 UT. The plot is dominated by low-frequency noise. Figure A3 does not show any other obvious periodicities at high frequencies, such as might be expected from quasi-periodic oscillations (see Chapter 10 of Hellier, 2001). There may be excess noise at log (frequency [cycles/day]) > 2.5, corresponding to periods of 4.6 minutes and shorter.

**3. Long-Term Variability**

Figure 3 is a plot of all the photometry of V849 Her that we took in 2010. Notice the fadings in magnitudes, during the middle and last nights. This behavior is similar to that of the VY Scl stars, which are nova-like variables that go into erratic low states. In these low states, the mass transfer rate through the accretion disk decreases, and in some cases almost completely shuts down (Robinson et al., 1981).

Figure 4 shows magnitudes observed by the American Association of Variable Star Observers (AAVSO) (Henden 2011), most of whom are amateur astronomers. The filled circles show the observed magnitudes, either in *V* for CCD observations, or in $m_{vis}$ for visual estimates made through the eyepiece. The open triangles are limits to observations: they are for when an eyepiece observer looked for V849 Her but couldn't see it. At these times, V849 Her was fainter than the triangles that are plotted. The AAVSO observations are too sparse to show any outburst behavior reliably, but there are no clear dwarf nova outbursts, aside from one point at MJD = 3015.043, a CCD observation of *V* =12.6. V849 Her may still have "stunted" outbursts, similar to those discovered by Honeycutt et al. (1998), which resemble dwarf nova outbursts of only 1-2 magnitudes' amplitude.

Figure 5 shows all the 2011 photometry of V849 Her from SRO. The shallow (0.5-magnitude) low states in Figure 5 were not as deep at those observed in several VY Scl stars by Honeycutt and Kafka (2004). A closeup of the deepest low state we observed is shown in Figure 6. They are unlike the low states of VY Scl stars in that they appear to recur quasi-periodically, over a cycle of about 12 days. We suspect this is only a quasi-periodicity because the low states at JD – 2455734 = 36, 45, and 58 shown in Figure 5 do not appear to repeat exactly, although gaps in the light curve complicate this.

Figure A4 in Appendix A is a log-log plot of the phase-dispersion minimization spectrum (Stellingwerf, 1978) for all the 2011 SRO photometry of V849 Her, which was plotted in Figure 5. Phase-dispersion minimization (pdm) is useful for finding periodicities that are not necessarily sinusoidal, unlike the Lomb-Scargle periodogram, which assumes that one sinusoid is present in Gaussian-distributed noise. Phase-dispersion minimization is a phase-folding technique: over a range of trial frequencies, the data are folded at each frequency and then the dispersion is calculated, with the frequency at which the minimum dispersion occurs taken to be the frequency of the periodicity. Figure A4 shows a peak at log *f* = 0 cycles/day from the diurnal cycle, along with its surrounding forest of aliases.

The strongest peak, at log $f$ = −1.297, has a frequency of 0.0505 cycles/day, corresponding to a period of 19.8 days. Since the fadings of V849 Her shown in Figures 3 and 5 appear to recur in about half this period, we conclude that the 2011 SRO photometry has insufficient frequency resolution to resolve the period, if any, unambiguously.

The AAVSO observations apparently do, however. Figure 7 is the AAVSO light curve of V849 Her from 2010 to 2011 October 30 UT (Henden, 2011). Figure A5 in Appendix A is a pdm spectrum (Stellingwerf, 1978) of the AAVSO photometry from 2009 to 2011 October 30 UT. The strongest peak is at a frequency of 0.08024 cycles/day, corresponding to a period of 12.462 ± 0.074 days. A Lomb-Scargle periodogram, shown in Figure A6 in Appendix A, gives an identical result, with a false-alarm probability of < $10^{12}$. This variation has an epoch (or time of average crossing, going from below average to above average) of JD 2454867.7470. Figure 8 is a close-up of the AAVSO light curve of V849 Her, showing 100 days during 2011. It clearly shows low states, occurring about every 11-14 days. Figure 9 is the AAVSO photometry of V849 Her from 2009 to 2011 October 30 UT, phase-folded over a 12.46-day period. It shows an amplitude of at least 0.090 magnitudes, although in a noisy curve, so that of this 12.46-day periodicity is real, it may only be a quasi-periodicity.

**4. Discussion**

Ak et al. (2008) used near-infrared magnitudes from the 2MASS survey (Cutri et al., 2003) to calculate a distance for V849 Her of 922 pc. This assumed *E(B-V)* = 0.073, which Ak et al. estimated from the position of V849 Her in the Galaxy. It also assumed an orbital period of 0.1409 days (3.382 hours), the minimum alias thought likely by Misselt and Shafter (1995). We will not attempt to improve this here, since the orbital period is still unknown, but it implies that $M_J$ = 5.06, which is consistent with V849 Her being a nova-like variable (Warner, 1987).

That V849 Her is a nova-like variable is suggested by all its published spectra (Ringwald, 1993; Munari and Zwitter, 1998; Rodríguez-Gil et al., 2007), its derived absolute magnitude, and its long-term variability. The photometric period of 3.394 ± 0.072 hours that we have measured may therefore be from permanent superhumps, either apsidal or nodal (see Chapter 6 of Hellier, 2001). Either would imply an orbital period between 3 and 4 hours, where nova-like variables are common (see Figure 5.16 of Hellier, 2001). Whether the 3.394-hour periodicity is from superhumps, though, or whether it is the orbital period or something else, will have to await measurement of the orbital period with a radial velocity study.

We also wish to encourage the AAVSO to continue monitoring V849 Her. We also encourage more thorough photometric monitoring, to show whether the 12.46-day variation is periodic, quasi-periodic, or neither. Rodríguez-Gil et al. (2007) noted that the optical spectrum of V849 Her resembles those of some nova-likes found by the Hamburg

Quasar Survey, including HS 0139+0559, HS 0229+8016, and HS 0642+5049 (Aungwerojwit et al., 2005). Might these have photometric behavior similar to V849 Her? We also encourage reanalysis of the light curves of VY Scl stars. Might any others also show quasi-periodicities?

**Acknowledgments**

This research used photometry taken at Fresno State's station at Sierra Remote Observatories. We thank Dr. Greg Morgan, Dr. Melvin Helm, and Dr. Keith Quattrocchi, and the other SRO observers for creating this fine facility. Other photometry was done at Mount Laguna Observatory, which is operated by the Department of Astronomy at San Diego State University. KSK was supported by the Research Experiences for Undergraduates program at San Diego State University, which was sponsored by the National Science Foundation, grant AST 08-50564. Spectra were obtained at MDM Observatory, which when the observations were taken was run by a consortium of the University of Michigan, Dartmouth College, and the Massachusetts Institute of Technology. This research has made use of the Simbad database, which is maintained by the Centre de Données astronomiques de Strasbourg, France. We acknowledge with thanks the variable star observations from the AAVSO International Database contributed by observers worldwide and used in this research.

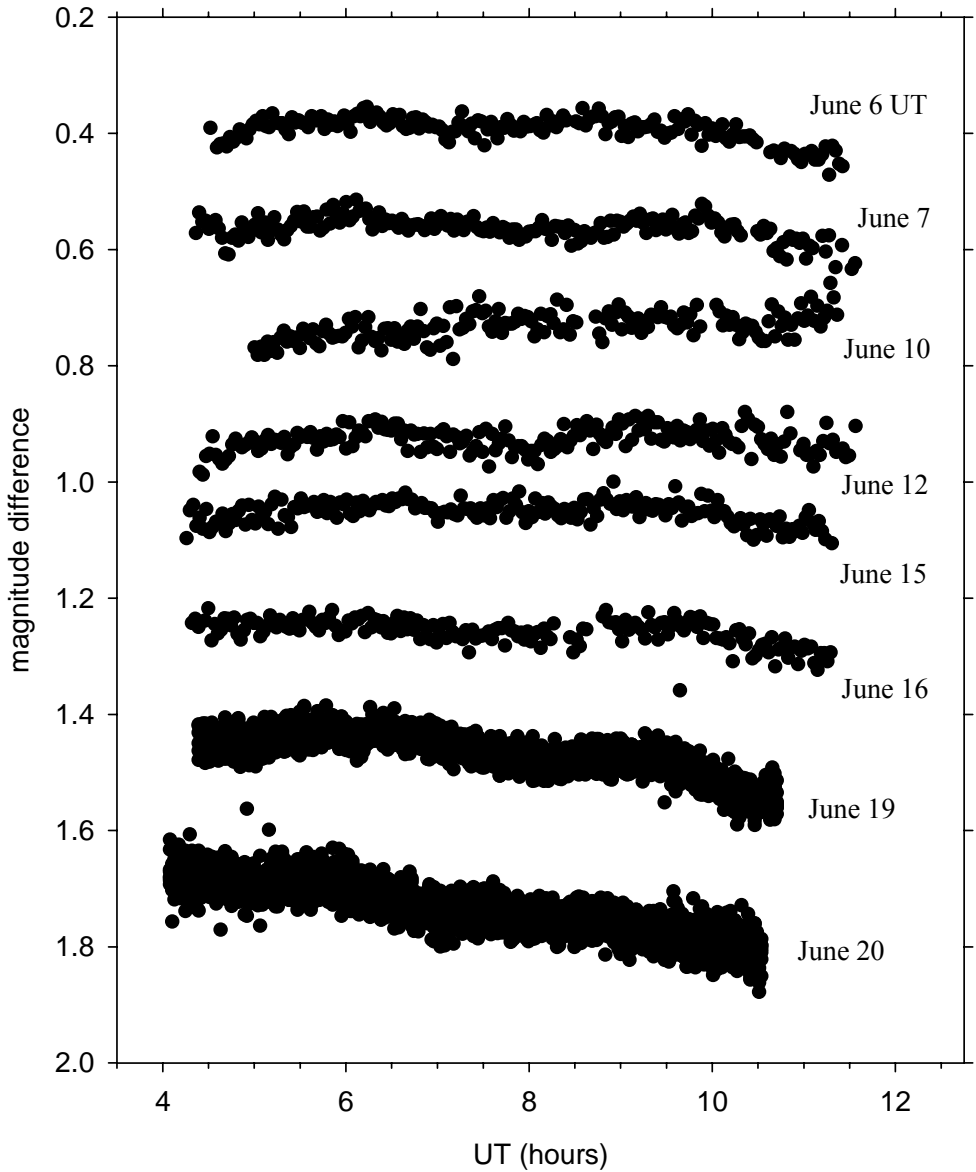

Figure 1.

Light curves of V849 Her, taken during 2010 June through a clear filter at SRO (for June 6-16 UT) and through no filter at Mount Laguna (for June 19 and 20 UT). To prevent these light curves from overlapping, each plot has been offset by 0, +0.1, −0.1, +0.4, +0.7, +0.9, +0.7, and +0.7 magnitudes, from top to bottom respectively.

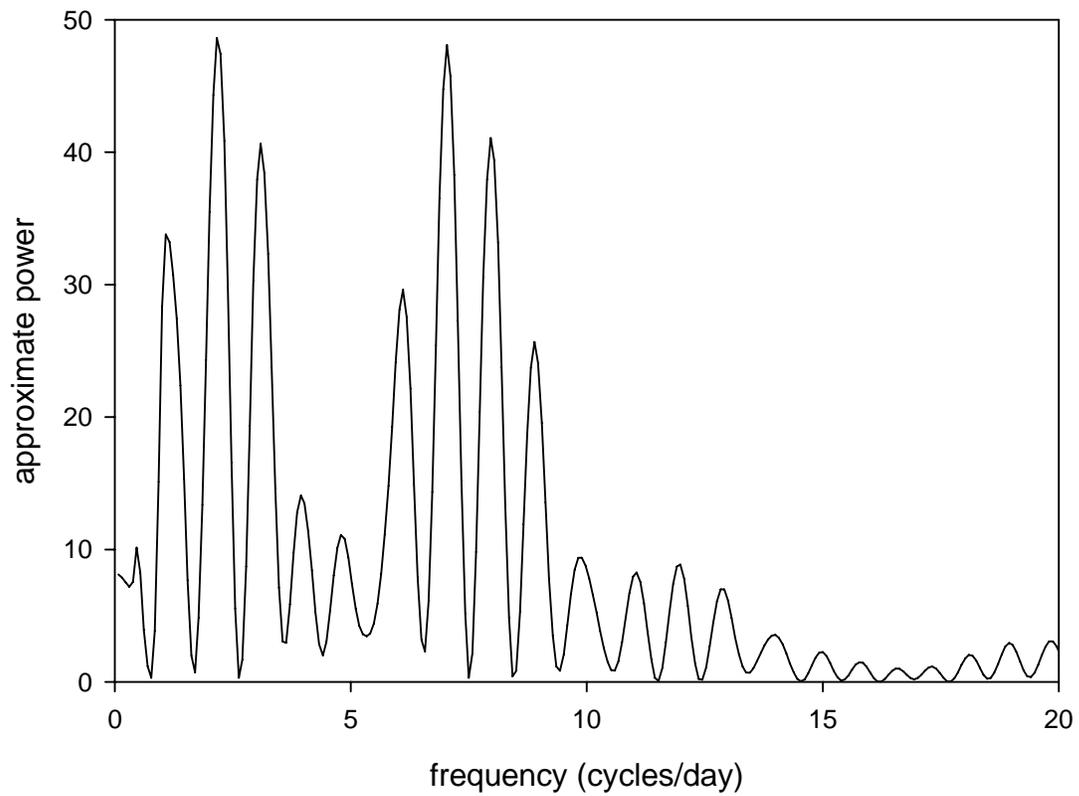

Figure 2.

A Lomb-Scargle periodogram for photometry taken at SRO during the last two nights (2010 June 15 and 16 UT). The peak at 7.073 ± 0.150 cycles/day corresponds to a period of 0.1414 ± 0.0030 days (3.394 ± 0.072 hours).

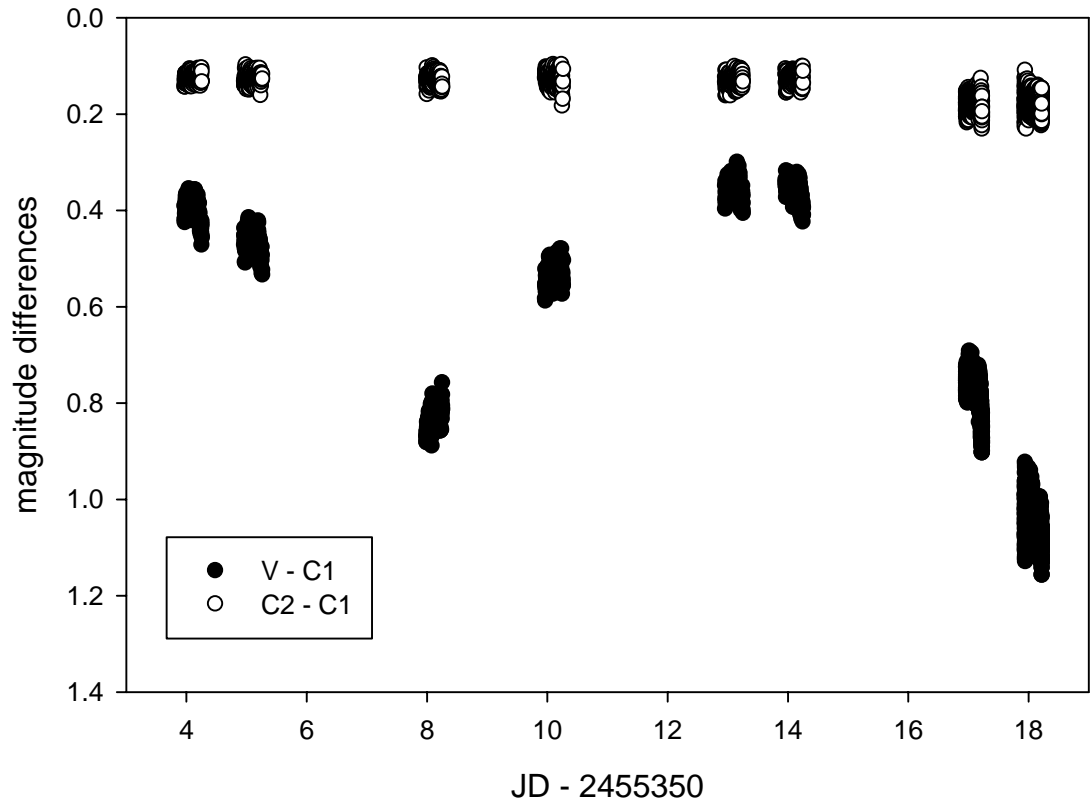

Figure 3.

The light curve of the 2010 June photometry of V849 Her, taken through a clear filter at SRO during the first six nights, and through no filter at Mount Laguna during the last two nights. Notice the fadings in magnitudes, during the middle and last nights. This behavior is similar to that of the VY Scl stars, a class of nova-likes that go into low states (see pages 171-173 of Hellier, 2001).

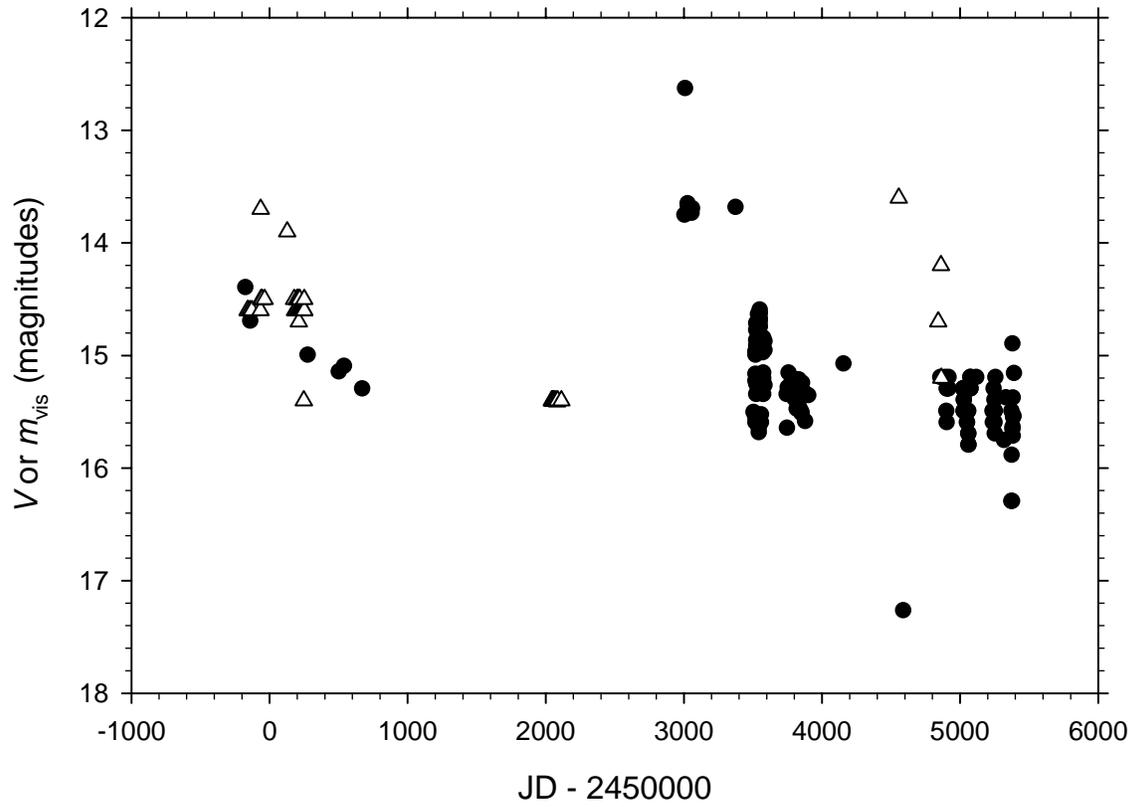

Figure 4.

The long-term light curve of V849 Her from AAVSO observations from 1995 to 2009 (Henden, 2011). Observed magnitudes, in either Johnson-Morgan $V$ or $m_{vis}$, are plotted with filled circles. The open triangles denote limits to observations: V849 Her was not detected because it was fainter than this, at these times.

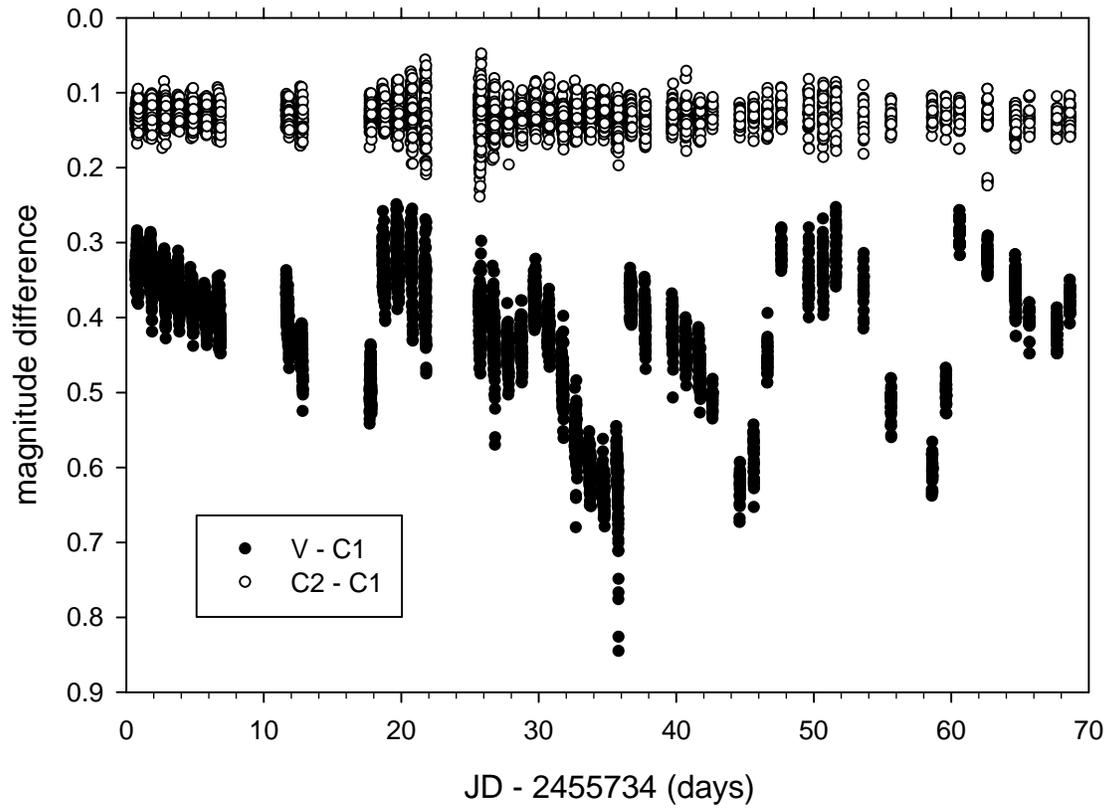

Figure 5.

The light curve for the 2011 photometry of V849 Her. It shows low states, similar to those of the VY Scl stars.

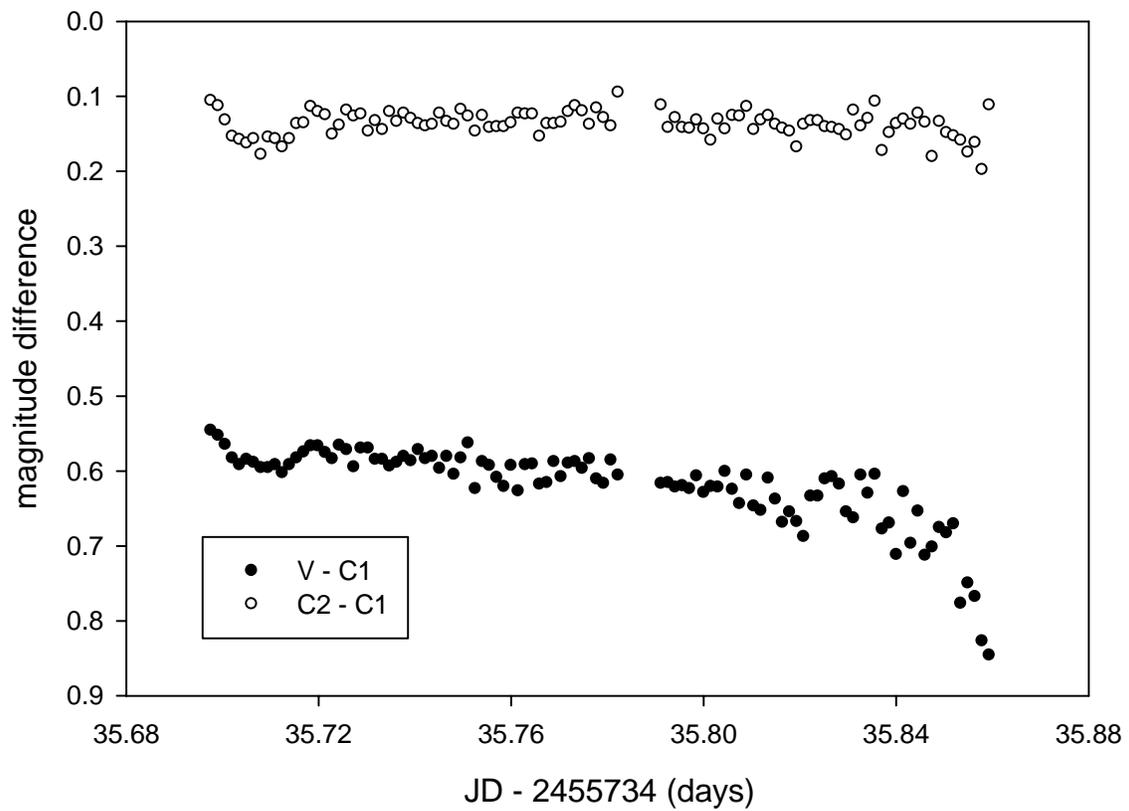

Figure 6.

The lowest state in the 2011 observations, on July 27 UT, has a shape like those observed in VY Scl stars by Honeycutt and Kafka (2004).

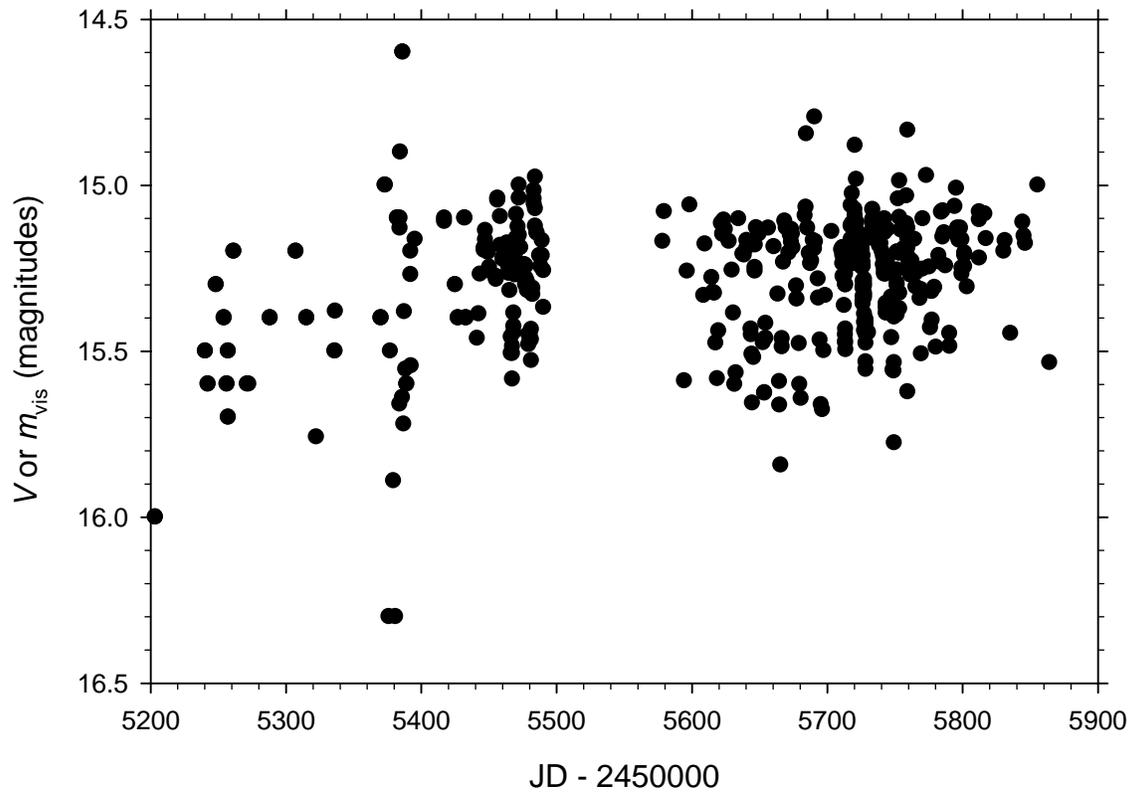

Figure 7.

The long-term light curve of V849 Her from AAVSO observations from 2010 to 2011 October 30 UT (Henden, 2011). Observed magnitudes, in either Johnson-Morgan *V* or $m_{vis}$, are plotted with filled circles.

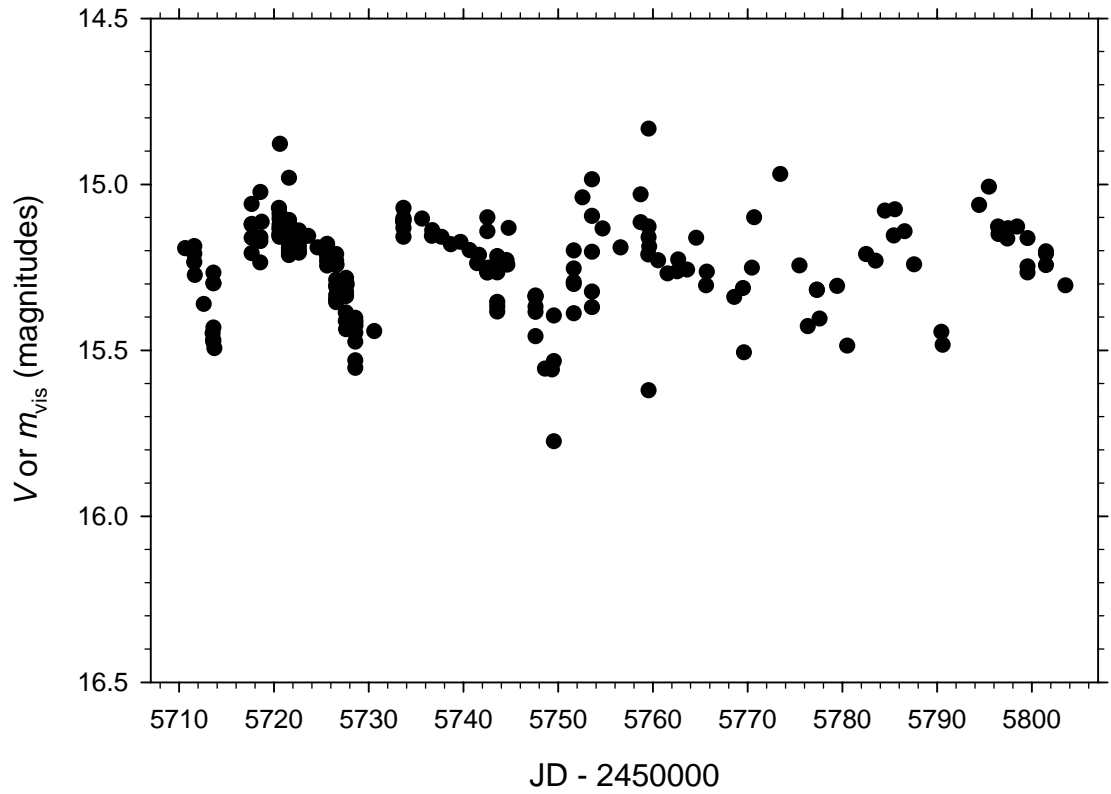

Figure 8.

A close-up of the light curve of V849 Her from AAVSO observations from 100 days in 2011. Observed magnitudes, in either Johnson-Morgan *V* or $m_{vis}$, are plotted with filled circles. Notice the low states, occurring approximately each 12.5 days.

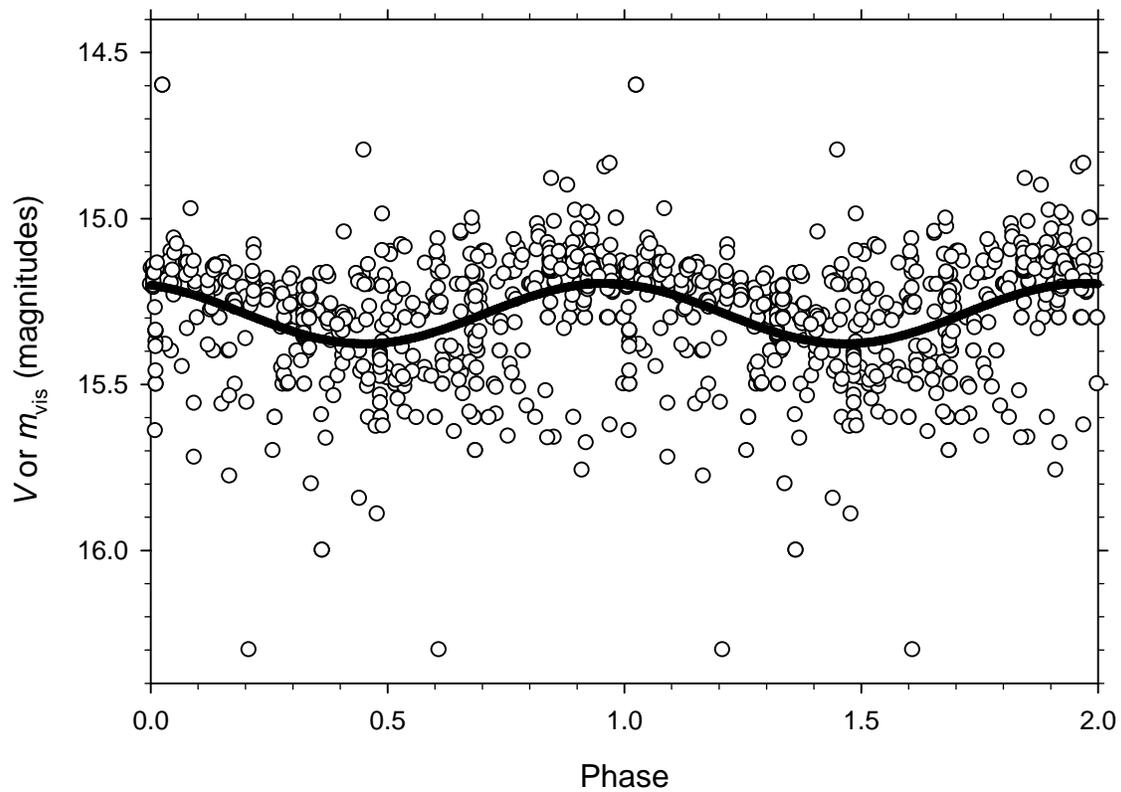

Figure 9.

The AAVSO photometry of V849 Her from 2010 to 2011 October 30 UT, phase folded on the 12.46-day period, and fitted to a sinusoid $V$ (or $m_{vis}$) = $y_0 + A \sin(\text{phase}/b)$, where $y_0 = 15.2865 \pm 0.0062$ magnitudes, $A = 0.0909 \pm 0.0080$ magnitudes, and $b = 1.013 \pm 0.029$ cycles.